\documentclass[11pt,twoside]{article}
\usepackage{asp2010}

\resetcounters
\aspvolume{455}
\aspcpryear{2012}
\aspvoltitle{4$^{th}$ {\em Hinode} Science Meeting: Unsolved 
Problems and Recent Insights}
\aspvolauthor{L.~R.~Bellot Rubio, F.~Reale, and M.~Carlsson, eds.}

\begin{document}

\title{The Effects of Including Non-Thermal Particles in Flare \\ Loop Models}
\author{K.~K.~Reeves,$^1$ H.~D.~Winter,$^1$ and N.~L.~Larson$^{1,2}$}
\affil{$^1$Harvard-Smithsonian Center for Astrophysics, USA}
\affil{$^2$Vanderbilt University, USA}

\begin{abstract}
 In this work, we use HyLoop \citep{2011ApJ...735..103W}, a loop model
 that can incorporate the effects of both MHD and non-thermal particle
 populations, to simulate soft X-ray emissions in various situations.
 First of all, we test the effect of acceleration location on the
 emission in several XRT filters by simulating a series of post flare
 loops with different injection points for the non-thermal particle
 beams.  We use an injection distribution peaked at the loop apex to
 represent a direct acceleration model, and an injection distribution
 peaked at the footpoints to represent the Alfv{\'e}n wave interaction
 model. We find that footpoint injection leads to several early peaks
 in the apex-to-footpoint emission ratio.  Second, we model a loop
 with cusp-shaped geometry based on the eruption model developed by
 \citet{2000JGR...105.2375L} and \citet{2005ApJ...630.1133R}, and find
 that early in the flare, emission in the loop footpoints is much
 brighter in the XRT filters if non-thermal particles are included in
 the calculation.  Finally, we employ a multi-loop flare model to
 simulate thermal emission and compare with a previous model where a
 semi-circular geometry was used \citep{2007ApJ...668.1210R}.  We
 compare the {\em Geostationary Operational Environmental Satellite}
 (GOES) emission from the two models and find that the cusp-shaped
 geometry leads to a smaller GOES class flare.
\end{abstract}

\section{Introduction}
Hard X-ray observations show that a large fraction (50--75\%) of the
energy liberated in solar flares can go into the generation of
non-thermal particles \citep[e.g.,][]{2005ApJ...626.1102S}. Thus these
particles are incredibly important for understanding the energy
transfer in solar flares.  However, many previous simulations of solar
flares do not include the effects of the non-thermal particles
directly \citep[e.g.,][]{1997ApJ...489..426H,
1998ApJ...500..492H,2002ApJ...578..590R,2006ApJ...637..522W,2007ApJ...668.1210R}.
In this work, we use the HyLoop suite of codes to simulate flares with
realistic geometries and we incorporate the effects of non-thermal
particles in order to further the current understanding of energy
transport in solar flares.  This suite of codes has been described in
detail previously in \citet{2011ApJ...735..103W}.

\section{HyLoop Model Description}

The HyLoop suite of codes consist of two codes that work together to
calculate the physical parameters in a loop due to MHD effects and
particle acceleration.  The MHD component of the codes is handled by
the Solar Hydrodynamic Equation Codes (SHrEC), which solves the
hydrodynamic equations along a loop in the following form:
\begin{eqnarray}
\frac{\partial n_{\rm e}}{\partial t} & = & -\frac{1}{a}\frac{\partial}{\partial s}\left(an_{\rm e}V\right), \label{eq:Simple_continuity}\\
\frac{\partial V}{\partial t} & = & \frac{-1}{n_{\rm e}m_{\rm p}}\frac{\partial P}{\partial s}+g_{\parallel}-V\partial V, \label{eq:Simple_momentum}\\
\frac{\partial\epsilon}{\partial t} & =& -\frac{1}{a}\frac{\partial}{\partial s}\left(a\epsilon V\right)-\frac{P}{a}\frac{\partial}{\partial s}\left(aV\right)-E_{\rm Rad} \nonumber \\
&& + \frac{1}{a}\frac{\partial}{\partial s}\left(aF_{\rm c}\right)+E_{\rm h}+E_{\rm NT},\label{eq:Simple_energy}\
\end{eqnarray}
where $g_{\parallel}(s)$ is the component of acceleration due to
gravity along the magnetic field, $P$ is the gas pressure, $a(s)$ is
the non-uniform cross-sectional area of the loop, $\epsilon(s,t)$ is
the internal energy density, $n_{\rm e}(s,t)$ is the electron density,
$\mathbf{V}(s,t)$ is the bulk plasma velocity, $F_{\rm c}$ is the
conductive flux, $E_{\rm Rad}$ is the radiative loss term, $E_{\rm h}$
is a background heating term, and $E_{\rm NT}(s,t)$, accounts for the
energy gained by the thermal plasma via collisions with the
non-thermal particle distribution.

The particles in HyLoop are handled by the Particle Tracking Codes
(PaTC), which models the evolution of the non-thermal particles using
direct Monte Carlo techniques.  In PaTC the non-thermal distribution
is treated as a series of test particles. These test particles are
randomly drawn from probability distributions designed to represent
the physics of a particular type of non-thermal beam.  PaTC accounts
for energy loss of the particles due to Coulomb collisions, and it
also accounts for the effects of a non-uniform magnetic field.  More
details about the HyLoop suite of codes and their assumptions,
boundary conditions, and initial conditions can be found in
\citet{2011ApJ...735..103W}.

\section{Effects of Spatial Location of Heating Using HyLoop}

It has long been assumed that non-thermal particles have been directly
accelerated by the current sheet formed during the flare, with the
particles being injected at the apex of post flare loops
\citep{1988ApJ...330L.131M,2006SoPh..236...59H}.  However, the number
of non-thermal electrons implied by hard X-ray observations has often
equaled or exceeded the total number of particles available in the
acceleration region. In order to solve this ``number problem'',
theories have been proposed that have particle acceleration occurring
in the denser chromosphere and transition
region. \citet{2008ApJ...675.1645F} have proposed a mechanism in which
the energy of a flare accelerates non-thermal particles in the
chromosphere via interactions with large-scale Alfv{\'e}n waves.

\begin{figure}[!t]
\centering
\includegraphics[scale=0.395]{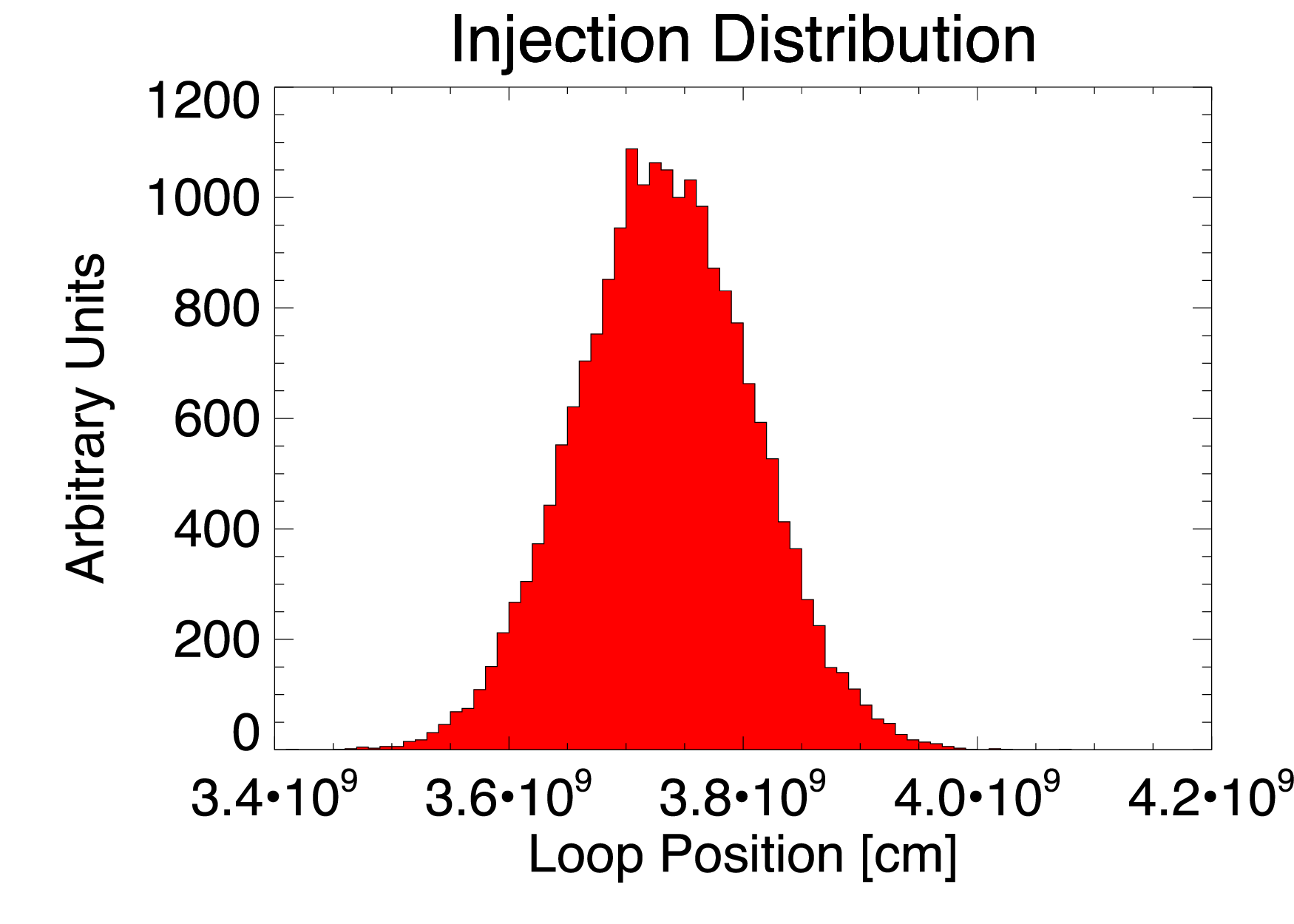}
\includegraphics[scale=0.395]{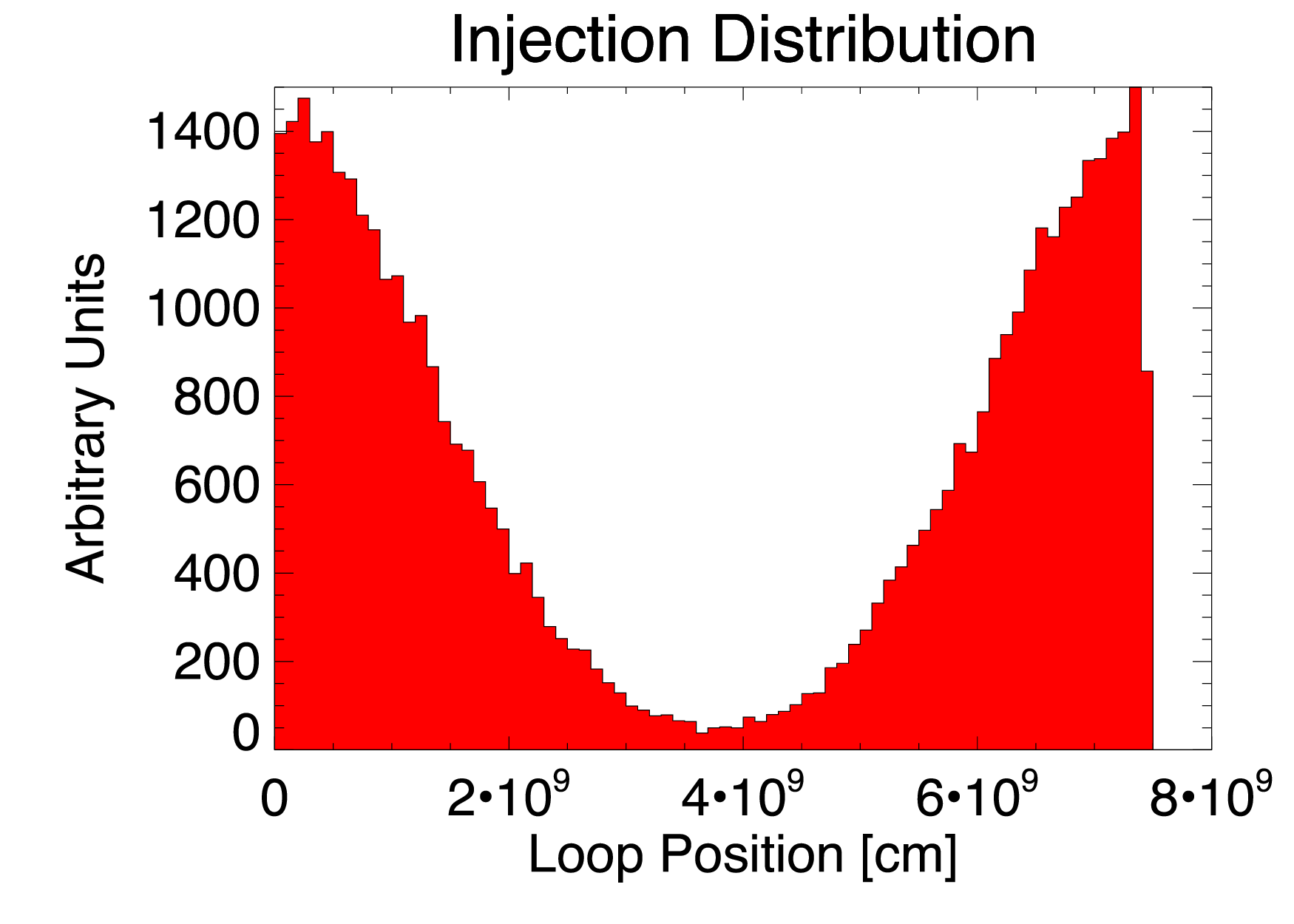}
\caption{Injection distributions for the two different experiments: 
non-thermal particle injection peaked at the loop apex ({\em left}), 
and non-thermal particle injection at the loop footpoints ({\em right}).  
\label{injection.fig}}
\end{figure}

In order to construct a simple model that addresses this proposal, we
use HyLoop to change the location of the non-thermal particle
injection and examine the results on the soft X-ray emission.  We
simulate two particle acceleration scenarios: the acceleration of
non-thermal particles at the loop apex (case a) and the acceleration
of non-thermal particles by interactions with large-scale Alfv{\'e}n
waves, which occurs preferentially at the base of the loops (case b).
We inject non-thermal particles into the loop using two different
distributions, one steeply peaked at the loop apex, and one steeply
peaked at the loop footpoints. The two distributions are shown in
Figure \ref{injection.fig}.

An estimated flare energy of 10$^{30}$~erg is put into the production
of non-thermal particles. These non-thermal particles are comprised
entirely of electrons.  The properties of the electrons are described
by probability distribution functions.  The pitch-angle probability
distribution function is characterized by a single parameter,
$\gamma$, which is set to $\gamma=0$ for both injection locations,
corresponding to non-thermal particle acceleration via stochastic
(turbulent) process at the loop apex \citep{2011ApJ...735..103W}.  At
the footpoints this pitch angle distribution also represents a
stochastic process used to represent particle interactions with
multiple wave-modes.  The energy probability for each simulation is
distributed as a single power-law function, $F(E)=E^{-\delta}$, with
$\delta=3$.

A pre-flare loop is heated using a heating function $E_{\rm h}(s)=H \,
T{^\alpha}P{^\beta}$ with $\alpha=3/2$ and $\beta=0$, which leads to
heating primarily at the loop apex.  The scaling constant, $H$, is
chosen to give the loop a $T_{\rm max}=1.5$~MK. The loop geometry is
defined by the Green (1965) current sheet model, and is the same
geometry used in \citet{2011ApJ...735..103W}.

\begin{figure}[p]
\includegraphics[scale=0.4]{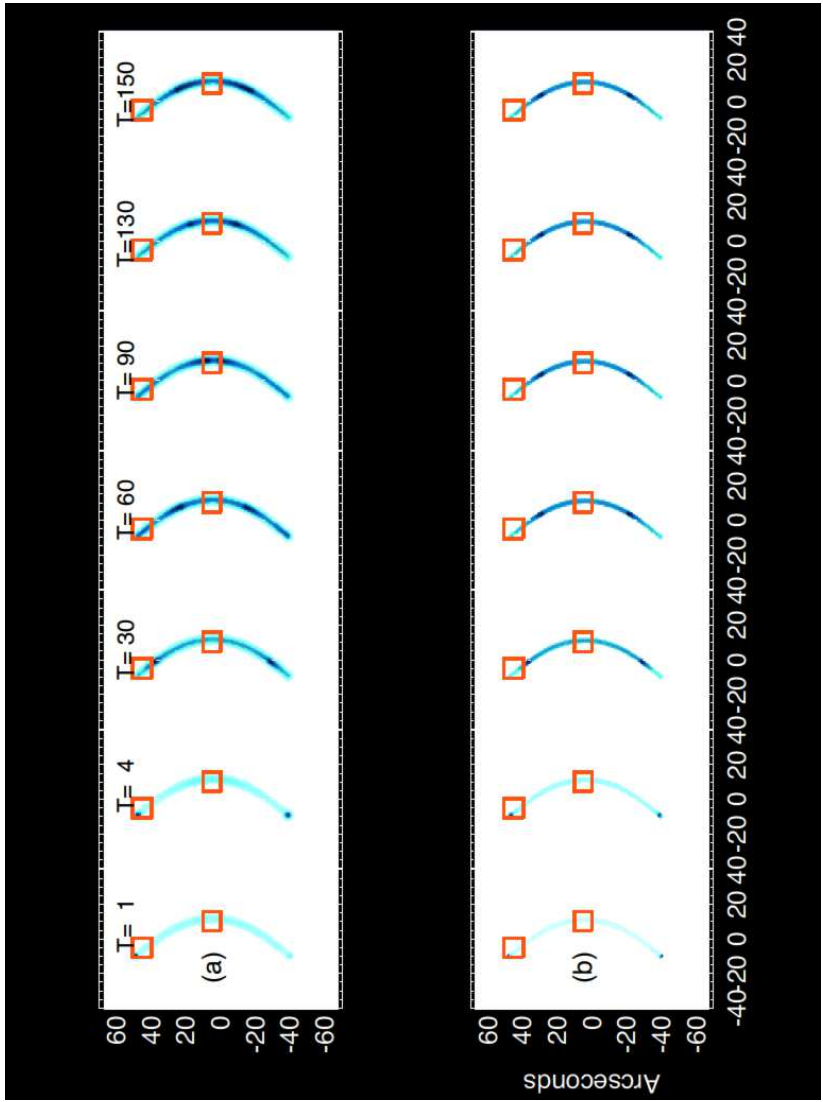}
\caption{ Simulated images for the XRT Ti-poly filter for {\em (a)} 
particle injection peaked at the apex, and {\em (b)} particle
injection peaked at the footpoints.  Cyan boxes indicate regions used
to calculate ratios. \label{ti_poly.fig}} \end{figure}

\begin{figure}[!t]
\centering
 \includegraphics[scale=0.65]{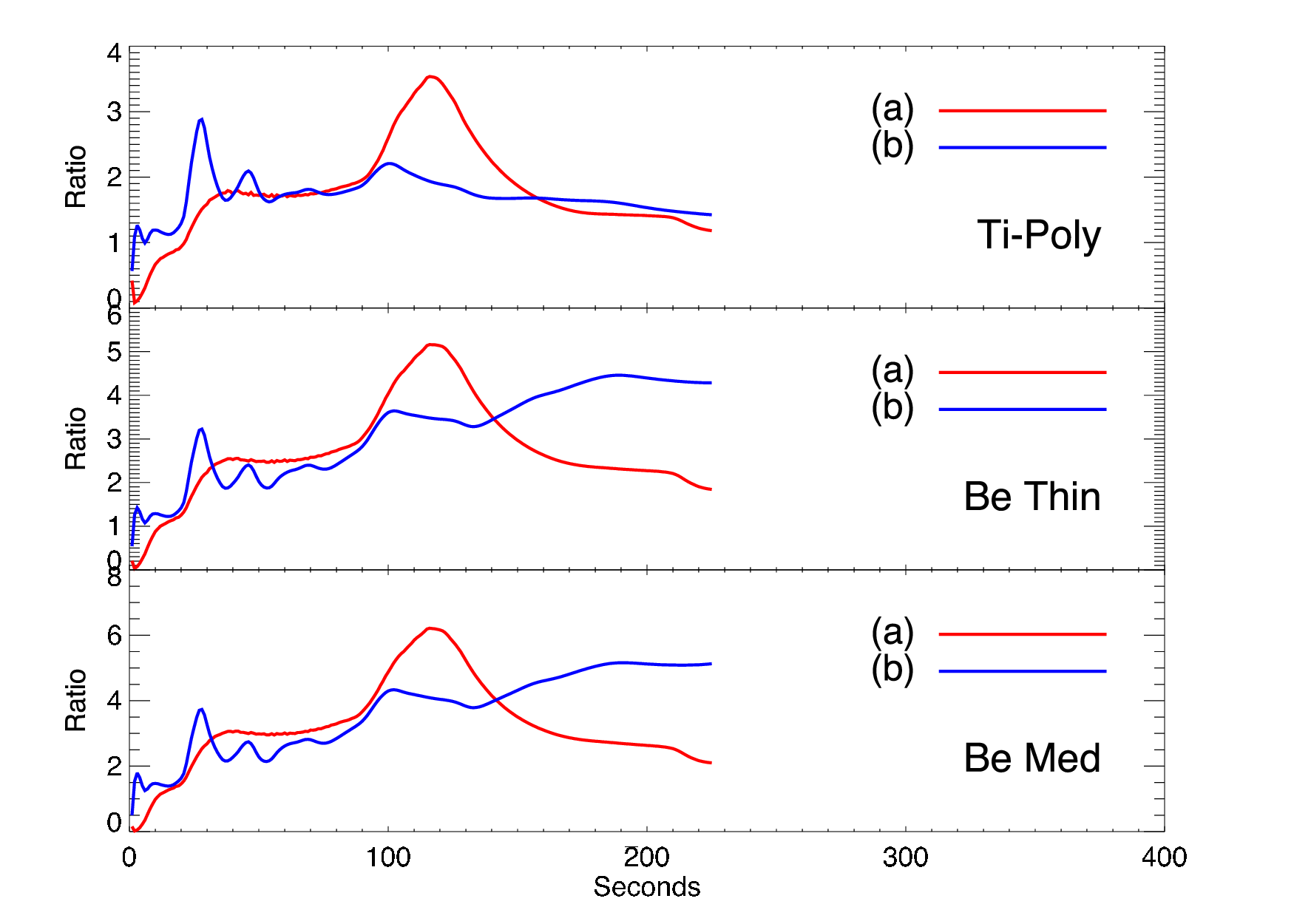}
\caption{Simulated apex-to-footpoint emission ratio for the 
Ti-poly ({\em top}), Be-thin ({\em middle}) and Be-med ({\em bottom})
filters.  Results from the simulation with the particle injection
peaked at the apex (case a) are shown in red and results from the
simulation with particle injection peaked at the footpoints (case b)
are shown in blue. \label{ratio.fig}} 
\end{figure}

Soft X-ray emission is simulated in three XRT filters, the Ti-poly
filter, the Be-thin filter and the Be-med filter.  Images of the
Ti-poly emission are shown in Fig.~\ref{ti_poly.fig} for both the
apex injection and the footpoint injection cases.  In both cases, the
footpoints show increased emission before the rest of the loop, and
there are bright knots of emission along the loop.  These bright knots
are locations of density enhancements in the loop due to chromospheric
evaporation.  The bright knots appear at different locations at each
time step in the two cases, indicating that that the timing of the
chromospheric evaporation is different depending on the location of
non-thermal particle injection.

We calculate the ratio between the apex and footpoint emission for the
Ti-poly, Be-thin, and Be-med filters for both particle injection
profiles.  The evolution of this ratio is shown in
Figure~\ref{ratio.fig}.  For the footpoint injection case, there is a
clear and sharp peak in the ratio at about 30~s after the onset of the
flare, followed by a secondary peak at about 45~s.  There is also a
broad peak in the ratio in all the filters at about 100~s, which is
when the apex density peaks in this case, increasing the apex
emission.  In the apex injection case, there is a steady increase in
the ratio starting at the flare onset, leading to a plateau between 
40--80~s, and then a large peak at about 120~s.  The large peak in the
ratio occurs at the same time as the peak apex density in this case.

The ratio plots of Fig.~\ref{ratio.fig} show that there are distinctly
different signatures in the soft X-ray emission for particle injection
at the apex versus at the footpoints.  These results could be used to
start looking for evidence of particle injection at the footpoints in
order to confirm the theory put forth by \citet{2008ApJ...675.1645F}.

\section{Cusped Flare Loop Geometry} 
Many flares have been observed to have a cusp-shaped geometry in the
soft X-rays \citep[e.g.,][]{1992PASJ...44L..63T, 1996ApJ...459..330F,
2008ApJ...675..868R}.  This geometry is defined by the reconnecting
magnetic fields that cause the release of energy in a solar flare.  In
the following section, we explore the effects of adding non-thermal
particles to a loop with a cusp-shaped geometry that is defined by a
model of reconnecting magnetic fields used to simulate coronal mass
ejections and flares.
 
\begin{figure}[!t]
\includegraphics[scale=0.45]{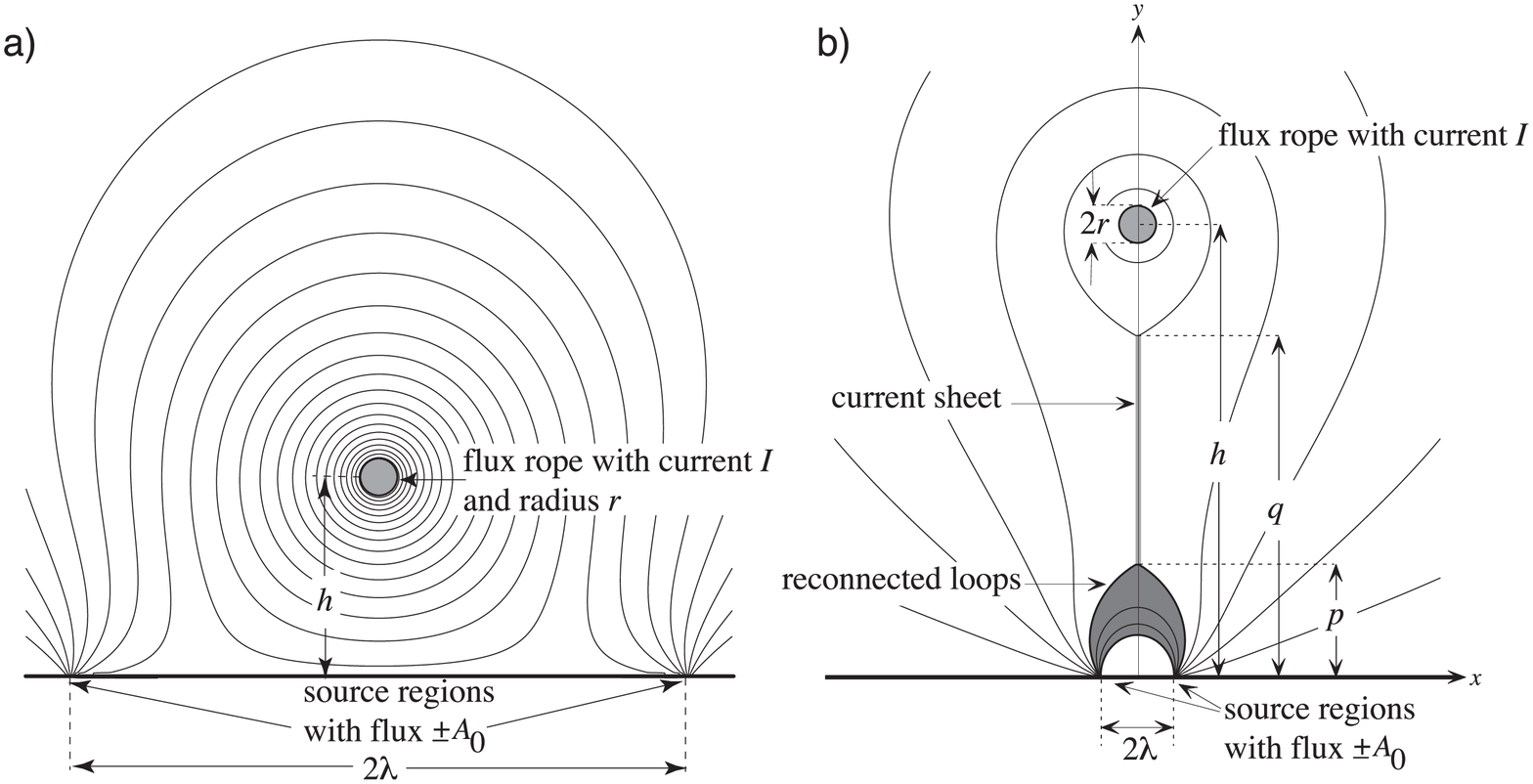}
\caption{{\em (a)} Magnetic configuration for the model prior to the 
loss of equilibrium. The current, $I$, is in the $z$ direction, out of
the plane of the figure.  {\em (b)} Magnetic configuration for the
model after the eruption and formation of the current sheet.  From
\citet{2006ApJ...644..592R}, reproduced by permission of the
AAS. \label{mag_config.fig}} \end{figure}
 
The model we use for the flare initiation is based on a version of the
\citet{2000JGR...105.2375L} model that has been expanded to include
gravity \citep{2004SoPh..219..169L, 2005IAUS..226..250R}.  In this
model, a flux rope is assumed to be in equilibrium prior to eruption
due to a balance among the gravitational force, the magnetic tension
force and the magnetic compression force.  There are two surface
sources that represent the sunspot magnetic field, and as these
surface sources are quasi-statically brought closer together, a loss
of equilibrium occurs and leads to an eruption.  After the initiation
of the eruption, a current sheet forms underneath the flux rope as
shown in Figure \ref{mag_config.fig}.  The current sheet is detached
from the solar surface, allowing an arcade of reconnected magnetic
loops to form between the localized sources as the eruption
progresses.

The physical parameters in the model for the case we consider here 
are as follows:
\begin{displaymath}
\begin{array}{lll}
M_{\rm A}  =  0.01, & &h_0 = 5 \times 10^{9} \:\mbox{cm}, \\
 m  = 2.1 \times 10^{16} \:\mbox{g},  & & \rho  =  1.67 \times 10^{-16}  
\:\mbox{g cm$^{-3}$}, \\
\ell =  10^{10} \:\mbox{cm}, & & B_0 = 25 \:\mbox{G},
\end{array}
\end{displaymath}
where $m$ is the mass of the flux rope, $\rho$ is the atmospheric
density at the base of the corona, $B_0$ is the background magnetic
field strength, $\ell$ is the length of the flux rope, and $h_0$ is
the height of the flux rope at the maximum current point on the
equilibrium curve---a convenient point to use for normalization
purposes \citep[see][]{2000JGR...105.2375L}.  $M_{\rm A}$ is the
inflow Alfv{\'e}n Mach number, which specifies the reconnection rate,
and it is fixed at the midpoint of the current sheet.  The parameters
listed above are chosen because they are the same as those used in
modeling one of the flares studied in \citet{2006ApJ...644..592R} and
\citet{2010ApJ...712..429R}.

\begin{figure}[!t]
\centering
\includegraphics[scale=0.75,bb=0 0 451 310]{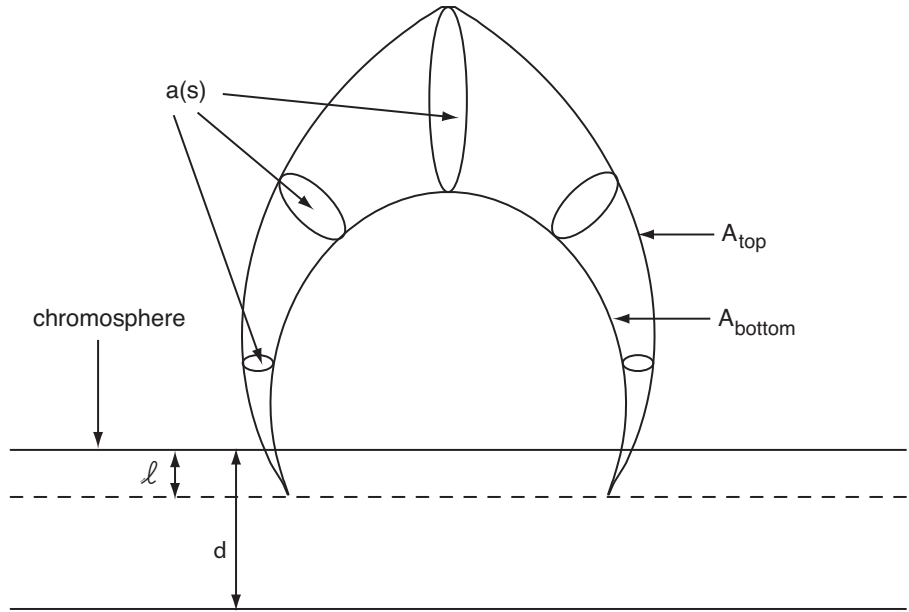}
\caption{The geometry of the flare loops in this model, as defined 
by the reconnecting magnetic fields.  The boundaries of the loops are
contours of the vector potential $A$ taken 20~s apart.  The cross
sectional area, $a(s)$, is defined by the width between the loop
boundaries.  A chromosphere is assumed to be at a height $\ell$ above
the bottom of the loop, to prevent the cross-sectional area from going
to zero.  \label{flare_loops.fig}} \end{figure}

We take as the boundaries of our loop contours of the vector potential
$A$ that travel through the reconnection region and form loops in the
\citet{2000JGR...105.2375L} reconnection model.  The top and bottom
boundaries are contours of the vector potential that travel through
the reconnection region 20~s apart.  This geometry is shown in
Figure~\ref{flare_loops.fig}.  The current version of the HyLoop code
cannot take into account the shrinkage of the loop, so we define the
loop geometry using a time late in the flare evolution, after the
shrinkage ceases to be important.  We then take the coordinates of the
loop and use them to define the loop geometry in HyLoop.

We assume that the cross section of the loop, $a(s)$, is a circular
cross section with the diameter equal to the width between the upper
and lower boundaries of the loop.  Because there are singularities in
the magnetic field at the footpoints of the loop, the cross-sectional
area would go to zero there (see Figure \ref{flare_loops.fig}).  So,
we assume that the chromosphere is a height $\ell=10^7$ cm above the
bottom of the loop, in order to avoid the singularities there.  We
also assume for a boundary condition that the chromosphere is held at
a temperature of $10^4$ K, and that it extends to a depth of
$d=2\times10^8$ cm.  This boundary condition is slightly different
from the one used in \citet{2011ApJ...735..103W}, but is commensurate
with boundary conditions used in other flare modeling codes
\citep[e.g.,][]{1987ApJ...319..465M}.

For this experiment, we input energy in two ways.  For the first case,
we input energy that is completely thermal in nature.  For the second
case, we input 100\% of the energy as non-thermal particles.  For both
cases, the total energy input into the loop is $1.35\times10^{29}$
erg, and it is input at the apex of the loop.  For the non-thermal
case, the energy is distributed as a single power-law function,
$F(E)=E^{-\delta}$, with $\delta=3$, as in the experiment described in
Section 2.  The pitch angle of the non-thermal particles is
distributed around zero, so that the velocities of the particles are
primarily aligned along the magnetic field.

\begin{figure}[p]
\includegraphics[scale=0.5]{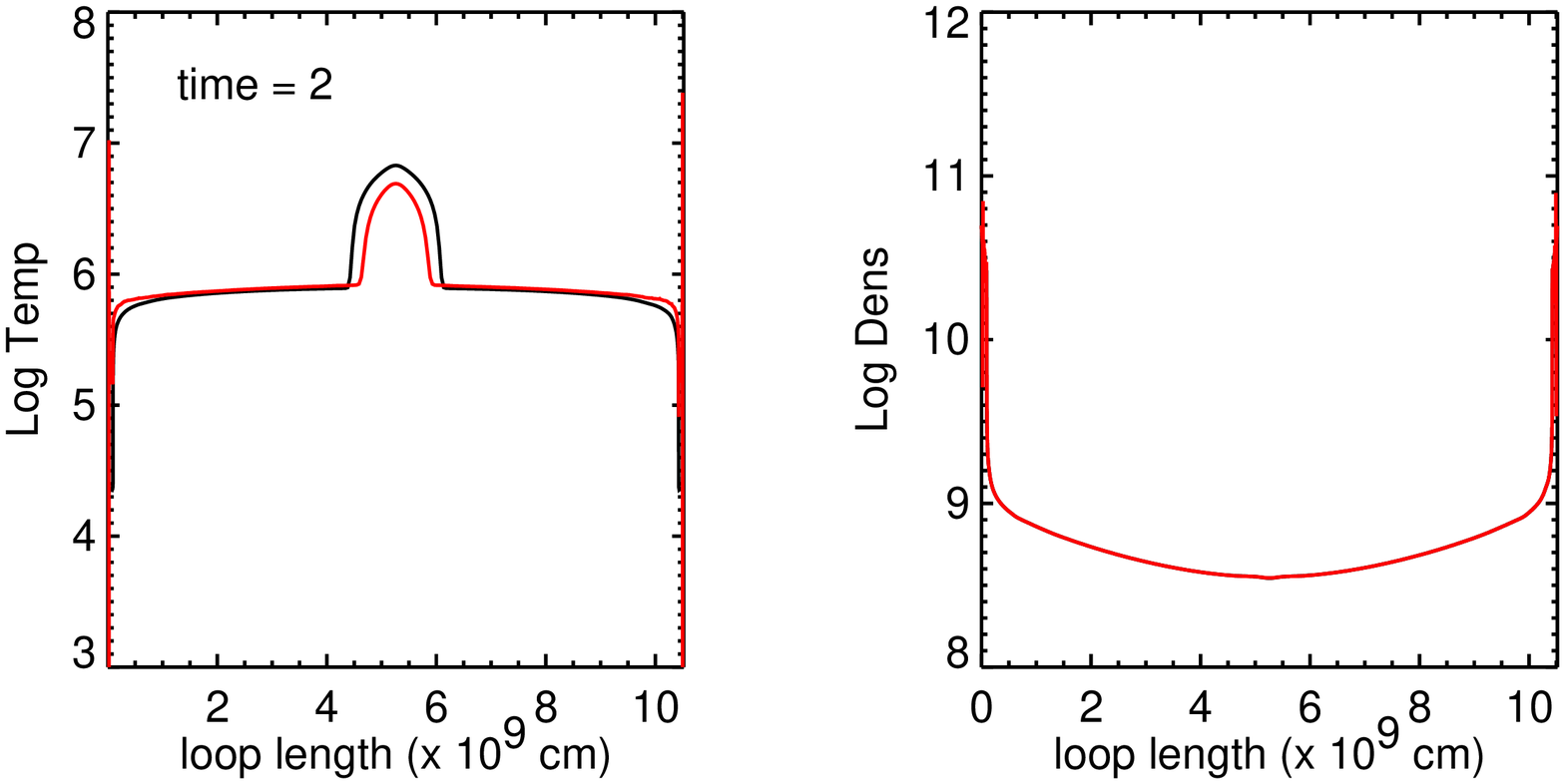}
\includegraphics[scale=0.5]{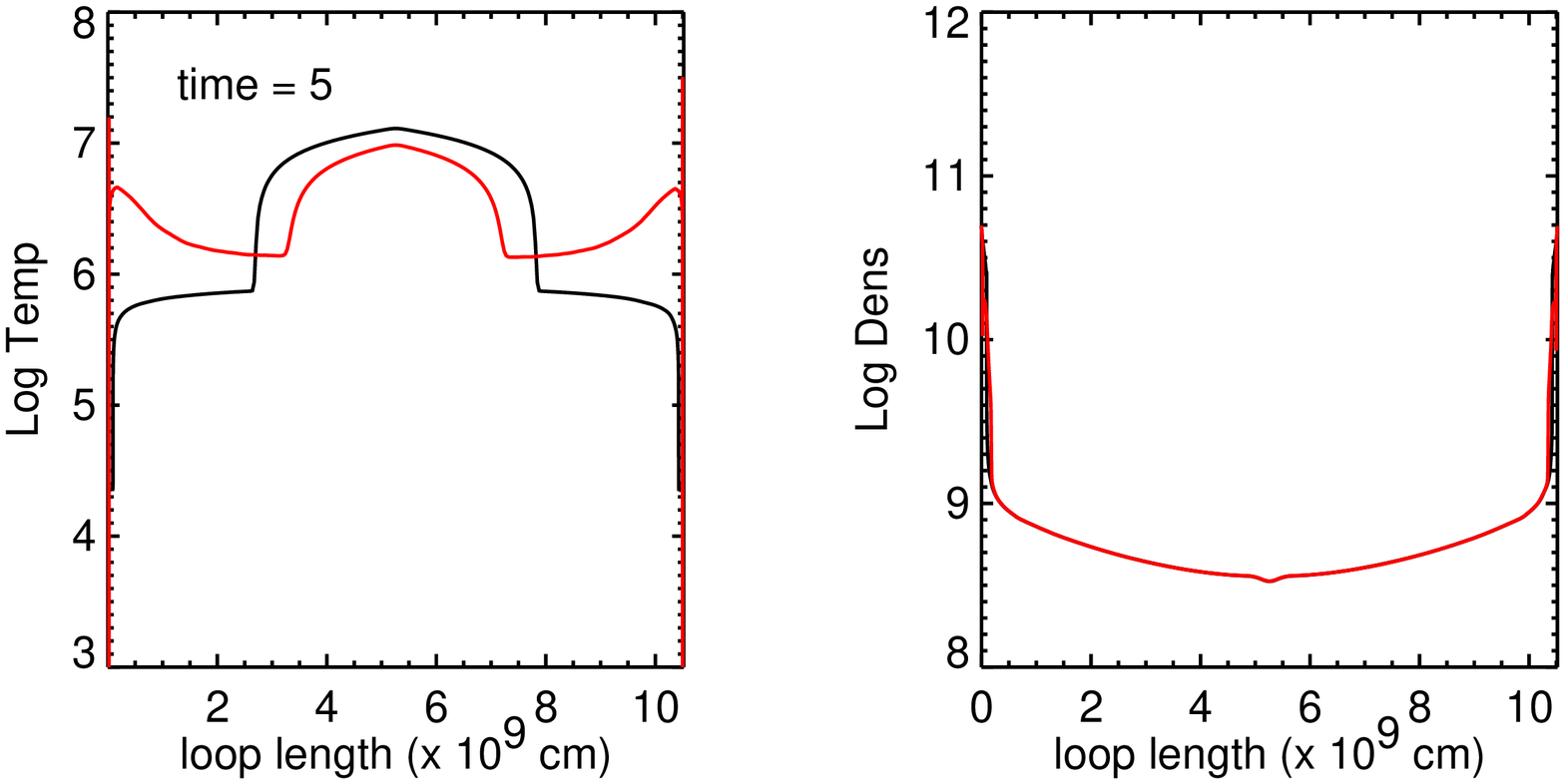}
\includegraphics[scale=0.5]{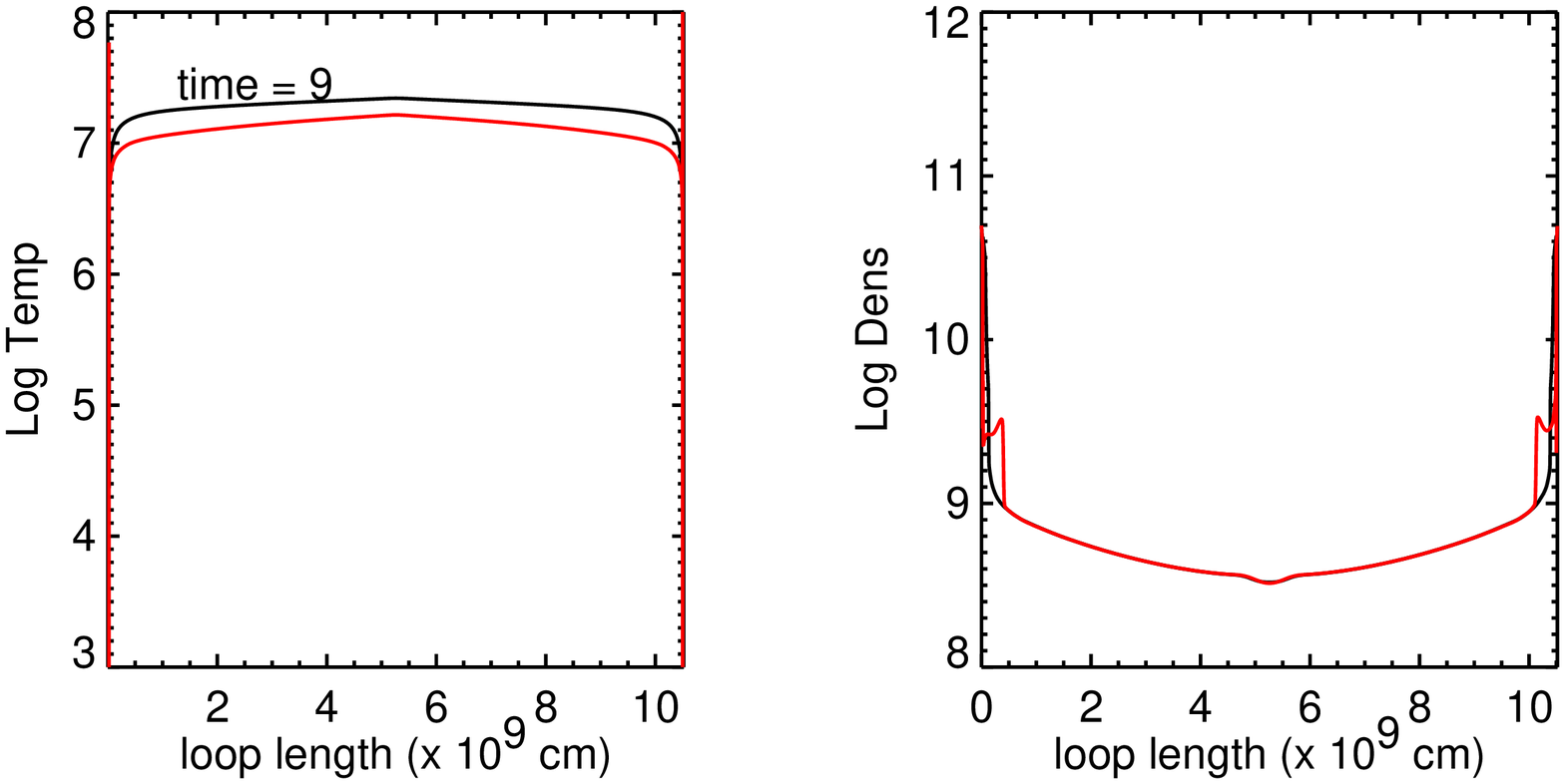}
\caption{Temperature ({\em left}) and density ({\em right}) for 
several different times during the beginning of the flare for a case
with no non-thermal particles (black) and a case where all of the
energy is put into non-thermal particles (red). Time is in
seconds. 
\label{diagnostics.fig}} 
\end{figure}

Figure~\ref{diagnostics.fig} shows the evolution of the temperature
and emission measure during the first few seconds of the flare.  For
the non-thermal case, plotted in red, the figure clearly shows that the
footpoints heat up more quickly than in the thermal case.  This
heating happens because the non-thermal particles stream down the
magnetic field lines within the first few seconds after initiation and
deposit their energy in the chromosphere, causing heating at the base
of the loop.

\begin{figure}[!t]
\includegraphics[scale=0.43]{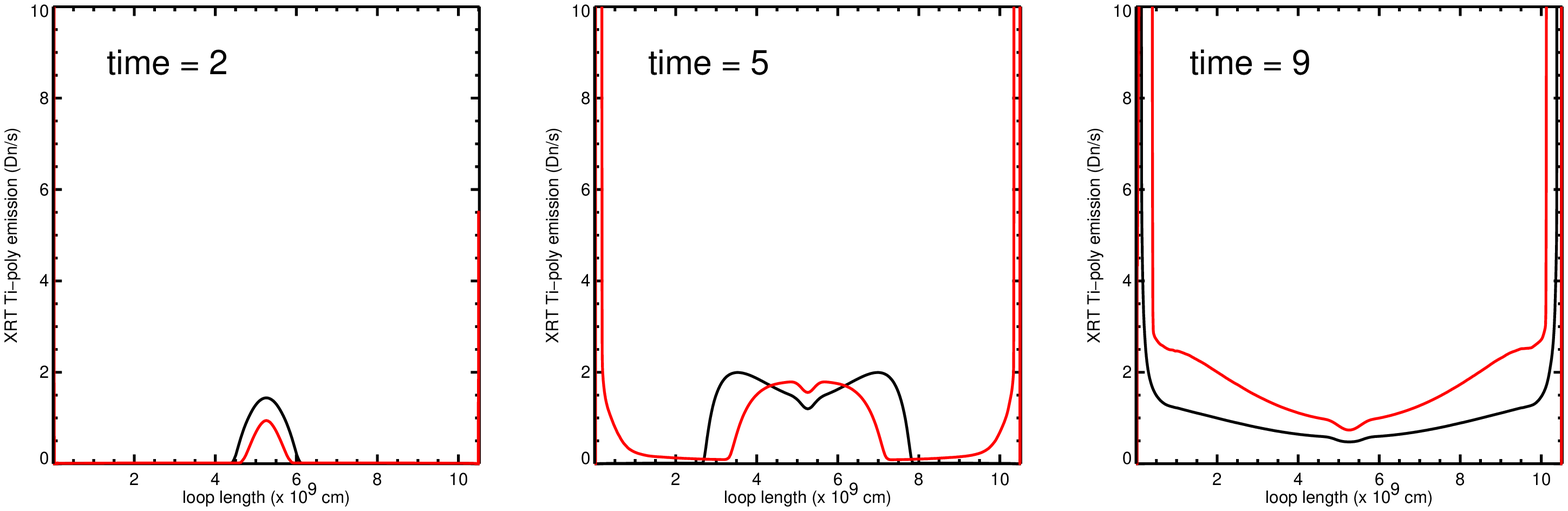}
\caption{XRT Ti-poly emission along the loop for several different 
times during the beginning of the flare for a case with no non-thermal
particles (black) and a case where all of the energy is put into
non-thermal particles (red). \label{xrt_emission.fig}} \end{figure}

\begin{figure}[!t]
\begin{center}
\includegraphics[scale=1,bb=-5 0 283 504]{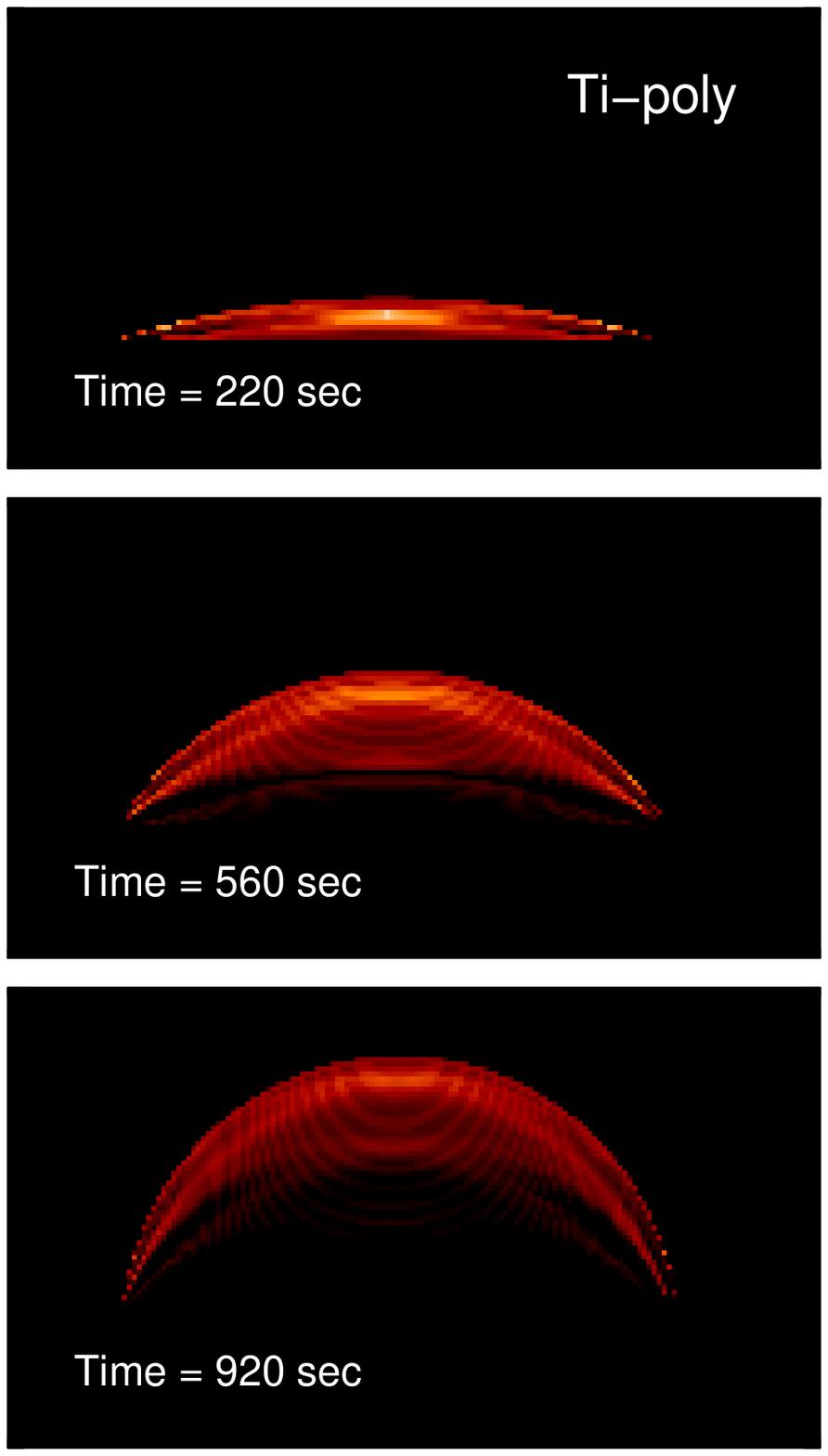}
\caption{XRT Ti-poly emission for a multi-stranded flare 
using the HyLoop code. \label{ti_poly_multistrand.fig}}
\end{center}
\end{figure}

Figure \ref{xrt_emission.fig} shows the XRT Ti-poly emission as a
function of loop length for the thermal and non-thermal energy
distributions for several times early in the flare evolution.  Both
cases show a peak in the loop-top emission early in the flare.
However, because of the elevated temperatures in the footpoints in the
loop with the purely non-thermal energy input, there is more footpoint
emission in that loop by 5 s than in the loop with only thermal energy
input.

\section{Multi-Stranded Flare}

The results presented above are interesting, but they are only
calculated for a single loop.  However, there is growing consensus in
the solar community that flares consist of many loops
\citep{1998ApJ...500..492H,2002ApJ...578..590R,2005ApJ...618L.157W, 
2007ApJ...668.1210R}.  Ultimately, we will use HyLoop to simulate a
multi-stranded flare that includes the effects of non-thermal
particles.  Here, we use HyLoop to simulate a multi-stranded flare
with completely thermal energy input, an experiment that takes less in
the way of computing resources than the full non-thermal case.  The
geometry in these loops is defined as in the previous section, with a
cusp-shaped top and tapered ends.

Using the same methodology as in \citet{2007ApJ...668.1210R}, we
simulate 140 discrete loops to make up the flare arcade, with a new
loop appearing every 20~s. The loss-of-equilibrium model parameters
are the same as in the previous section.  The input heating rate in
each loop is given by
\begin{equation}
\epsilon(s,t) = \epsilon_0 + f(s)\, g(t) \, \epsilon_{\rm flare},
\end{equation}
where $\epsilon_0$ is the background heating parameter, $g(t)$ is a
triangular function, and $f(s)$ is a Gaussian function given by
$\exp[-(s -s_0)^2/2\sigma^2]$, where $s_0$ is the apex of the loop,
$\sigma$ is the width of the heating region (taken to be $10^8$~cm),
and $\epsilon_{\rm flare}$ is the constant amplitude of the heating.  The
total energy input into each loop is calculated from the Poynting flux
into the current sheet in the loss-of-equilibrium calculation, as in
\citet{2007ApJ...668.1210R}.

\begin{figure}[t]
\includegraphics[scale=.5]{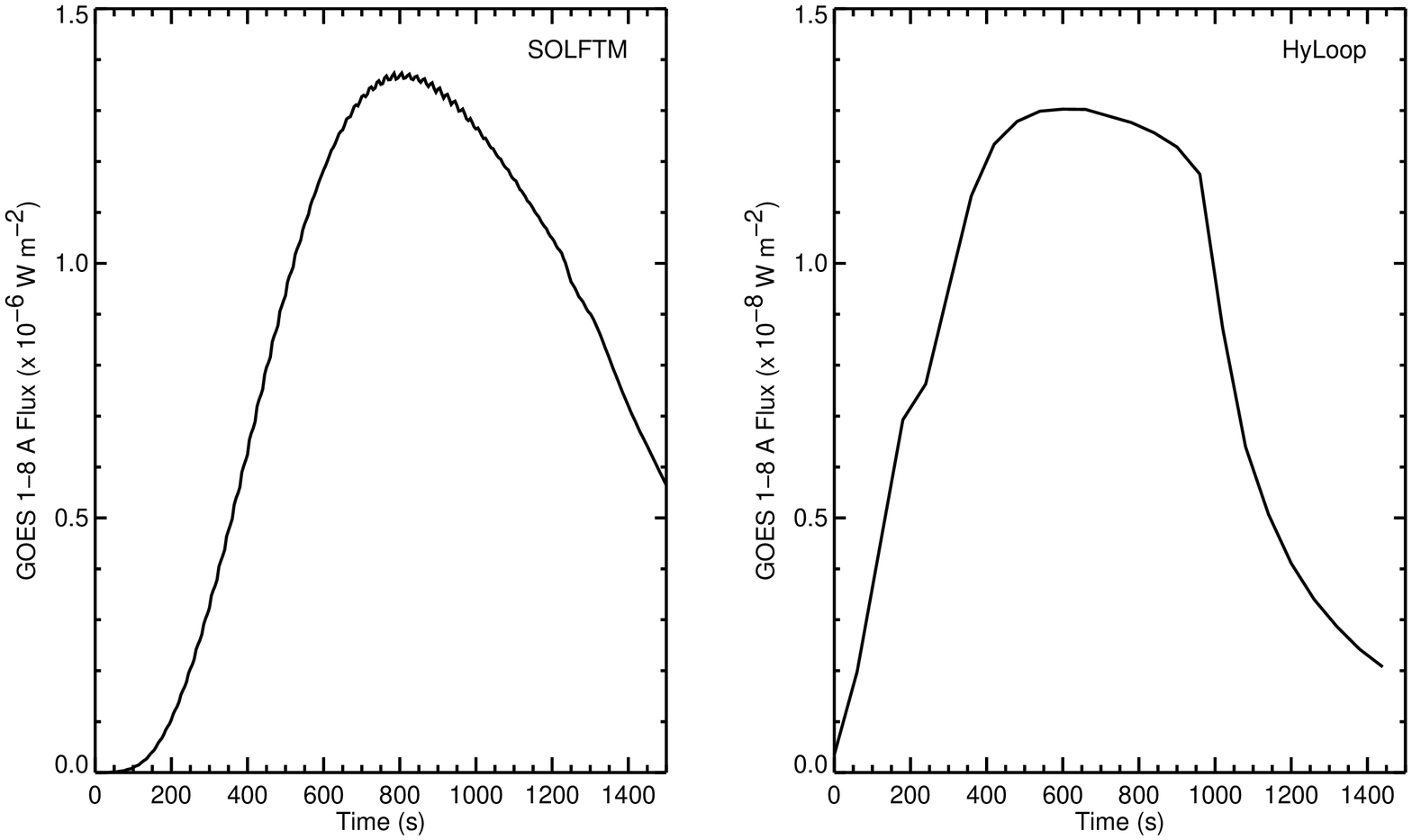}
\caption{GOES 1--8 \AA\ light curves for flares 
simulated using HyLoop with a cusp-shaped geometry ({\em right}) and
SOLFTM with a semicircular geometry ({\em
left}). \label{compare_goes.fig}} \end{figure}
 
Using this energy input, we simulate a multi-stranded flare and the
XRT Ti-poly intensity.  Figure~\ref{ti_poly_multistrand.fig} shows the
evolution of the simulated Ti-poly emission for the multi-stranded
flare.  The effects of the geometry can be clearly seen in this
figure, since the ends of the loops are very tapered as compared to
the tops of the loops.  In the first two frames, a compact brightening
can be seen at the top of the flare arcade.  This brightening is
caused by density fronts colliding at the tops of the loops to create
enhanced emission measure there.  This effect was found previously
with semi-circular loops by
\citet{2007ApJ...668.1210R}.  This loop-top source persists because
new loops are constantly forming and producing these colliding density
fronts as the reconnection progresses.  Late in the flare, at $t=920$
s, there is still a small enhancement in the emission at the flare
loop top, but is is not as strong as early in the flare, since the
loops that are reconnected late in the flare have a lower volumetric
heating rate than those energized early in the flare.

In order to understand the effects the cusp-shaped geometry has on the
simulated flare, we compare two different models: HyLoop, with the
cusp-shaped geometry described in the previous section, and the NRL
Solar Flux Tube Model \citep[SOLFTM; see][]{1987ApJ...319..465M} with
semi-circular loops.  We model the GOES flux in each simulated flare,
as shown in Figure \ref{compare_goes.fig}.  We find that the HyLoop
model with the cusp-shaped geometry results in much weaker flare X-ray
emission than the SOLFTM model with the semi-circular geometry.  The
cusp-shaped geometry results in the weaker flare because there is a
smaller area in the loop footpoint for the heat to be deposited,
resulting in smaller evaporation flows and less chromospheric
evaporation in the cusp-shaped loop.

\section{Conclusions}

We have performed several different experiments using the HyLoop code
and found that non-thermal particles can have a profound effect on the
soft X-ray signal from a flare.  We have shown that the existence of
non-thermal particles in a beamed distribution will cause the
footpoints of the flare to heat up faster than if no non-thermal
particles are present.  The location of the input of the non-thermal
particles will have an effect on the evolution of the
apex-to-footpoint ratio, with non-thermal particles input at the
footpoints causing many small peaks in the ratio that are not seen if
the particles are input in the loop-top.  This information may be used
in conjunction with observational signatures to confirm or deny the
injection of non-thermal particles at the footpoints.

We have also investigated the effects of simulating a flare with a cusp-shaped geometry taken from the magnetic reconnection model of \cite{2000JGR...105.2375L}.  We find that emission in the loop footpoints is much brighter in the XRT filters if non-thermal particles are included in the calculation because the beamed particles quickly travel to the chromosphere and deposit their energy there.  We also find that the cusp shape of the loops causes a bottleneck in the loop that causes a smaller upflow of density than if a semi-circular loop is used.  The smaller upflow leads to lower densities in the loop, and thus a lower GOES class for the flare.

In the future, we plan to model multi-threaded flare arcades using HyLoop to incorporate the effects of non-thermal particles in each of the individual strands.  

\acknowledgements 
\begin{sloppypar}
{\em Hinode} is a Japanese mission developed and launched by 
ISAS/JAXA, with NAOJ as domestic partner and NASA and STFC (UK) as
international partners. It is operated by these agencies in
cooperation with ESA and NSC (Norway). KKR and HDW are supported by
contract NNM07AB07C from NASA to SAO.  Additional support for KKR
comes from the NSF-SHINE program, grant number ATM-0752257.
Additional support for HDW comes from NASA grant NNX09AB18G-R.  The
work of NLL was supported by the NSF-REU solar physics program at CfA,
grant number ATM-0851866.
\end{sloppypar}

\bibliography{hinode4}

\end{document}